\documentclass[a4paper, 11pt]{article}

\usepackage{graphicx}
\usepackage{amsmath}
\usepackage{amsfonts}
\usepackage{amssymb}
\usepackage{authblk}
\usepackage{tikz}
\usetikzlibrary{arrows,backgrounds,snakes,patterns}
\usetikzlibrary{shapes,arrows,chains}
\usepackage{subcaption} 
\usepackage{arydshln} 
\usepackage{verbatim}
\usepackage{booktabs}
\usepackage[margin=0.85in]{geometry}
\usepackage{multirow}
\usepackage{changes} 
\usepackage[flushleft]{threeparttable}

\begin{document}

\title{EnzyNet: enzyme classification using 3D convolutional neural networks on spatial representation}

\author[1,2]{Afshine Amidi}
\author[2]{Shervine Amidi}
\author[3]{Dimitrios Vlachakis}
\author[3]{Vasileios Megalooikonomou}
\author[2]{Nikos Paragios}
\author[2,3]{Evangelia I. Zacharaki}

\affil[1]{Massachusetts Institute of Technology, Cambridge, United States of America}
\affil[2]{Center for Visual Computing, Department of Applied Mathematics, \'Ecole centrale Paris, CentraleSup\'elec, Ch\^atenay-Malabry, France}
\affil[3]{MDAKM Group, Department of Computer Engineering and Informatics, University of Patras, Patras, Greece}

\maketitle

\begin{abstract}
During the past decade, with the significant progress of computational power as well as ever-rising data availability, deep learning techniques became increasingly popular due to their excellent performance on computer vision problems.
The size of the Protein Data Bank has increased more than 15 fold since 1999, which enabled the expansion of models that aim at predicting enzymatic function via their amino acid composition. Amino acid sequence however is less conserved in nature than protein structure and therefore considered a less reliable predictor of protein function.
This paper presents EnzyNet, a novel 3D-convolutional neural networks classifier that predicts the Enzyme Commission number of enzymes based only on their voxel-based spatial structure. The spatial distribution of biochemical properties was also examined as complementary information.
The 2-layer architecture was investigated on a large dataset of 63,558 enzymes from the Protein Data Bank and achieved an accuracy of 78.4\% by exploiting only the binary representation of the protein shape. Code and datasets are available at \texttt{https://github.com/shervinea/enzynet}.
\end{abstract}

\section*{Introduction}
The exponential growth of the number of enzymes registered in the Protein Data Bank urges a need to propose a fast and reliable procedure to classify every new entry into one of the six standardized enzyme classes denoted by the Enzyme Classification number (EC): Oxidoreductase (EC1), Transferase (EC2), Hydrolase (EC3), Lyase (EC4), Isomerase (EC5), and Ligase (EC6). Previous work included the use of traditional machine learning techniques requiring a tedious and crucial feature extraction step based on amino-acid sequence alignment \cite{Kumar2012} or using structural descriptors without relying on sequence alignment \cite{Dobson2005}.
A limitation in the aforementioned approaches is that the features have to be predefined, the appropriate choice of features affects the prediction accuracy and there is limited flexibility for model changes or updates (all preprocessing steps have to be repeated). These drawbacks are overcome by deep learning techniques that take care of extracting the relevant features seamlessly from the input.

With the common availability of data and an ever-increasing computing power, deep learning approaches, such as convolutional neural networks (CNNs), proved to be very efficient and outperformed traditional approaches. While CNNs have been used to predict protein properties in \cite{Lin2016}, Zacharaki \cite{Zacharaki2017} proposed a 2D CNN ensemble to classify enzymes and achieved 90.1\% accuracy on a benchmark of 44,661 enzymes from the PDB database. The protein structure was represented by multiple 2D feature maps characterizing the backbone conformation (torsion angles) and the (pairwise) amino acids distances. The results showed that the purely structural features (torsion angles) had limited contribution. This may be related to their global nature coming from the 2D representation, which fails to characterize the local 3D shape.

Recently, architectures directly dealing with 3D structures were tested on various datasets. Maturana et al. \cite{Maturana2015} presented a 3D convolutional neural network approach and benchmarked it on traditional datasets (LiDAR, RGBD, CAD) beating traditional approaches. Hegde et al. \cite{Hegde2016} presented two different representations of the 3D data representation: one via voxels, and the other one via projected 2D pixel images.

\section*{Materials and methods}
\subsection*{Dataset}

The dataset used in this study has been retrieved from the RCSB Protein Data Bank\footnote{Website (\texttt{http://www.rcsb.org/})}, which contained 63,558 enzymes in mid-March of 2017. It has been randomly split in training and testing set with the proportions 80\%/20\%. Also, 20\% of the training set has been put aside for validation and was later used for model selection. The details of each of those sets have been summed up in Table~\ref{table:dataset}.

\renewcommand{\arraystretch}{1.3}
\setlength{\tabcolsep}{5.0pt}
\begin{table}[!h]
\begin{center}
\begin{tabular}{lcccccccl}
\toprule
\textbf{EC1} & \textbf{EC2} & \textbf{EC3} & \textbf{EC4} & \textbf{EC5} & \textbf{EC6} & \textbf{Total} & \\
7,096 &12,081 & 15,290 & 2,875 & 1,703 & 1,632 & 40,677 & training \\
1,775 & 2,935 & 3,809 & 743 & 488 & 419 & 10,169 & validation \\
2,323 & 3,717 & 4,762 & 858 & 571 & 481 & 12,712 & testing \\
\cdashline{1-9}
11,194 & 18,733 & 23,861 & 4,476 & 2,762 & 2,532 & 63,558 & total \\
\bottomrule
\end{tabular}
\end{center}
\caption{Structure of the dataset}
\label{table:dataset}
\end{table}

\subsection*{Occupancy grid}
This paper aims at building a model that seamlessly extracts relevant shape features from raw 3D structures. 

For this purpose, enzymes are represented as a binary volumetric shape with volume elements (voxels) fitted in a cube $V$ of a fixed grid size $l$ with respect to the three dimensions. Continuity between the voxels is achieved by nearest neighbor interpolation, such that for $(i,j,k) \in [\![0;l-1]\!]^3$ a voxel of vertices 
\begin{equation*}
(i + \delta x, \, j+ \delta y, \, k+ \delta z) \quad | \quad (\delta x, \delta y, \delta z) \in \{0,1\}^3 
\end{equation*}

\noindent takes the value 1 if the backbone of the enzyme passes through the voxel, and 0 otherwise. 

To construct this shape representation, some preprocessing steps are necessary. First, protein structure is mapped to a grid of predefined resolution. The selection of grid resolution determines the level of complexity/scale retained for the enzymatic structures. A full resolution is not preferred due to high data dimensionality and because fine local details are less relevant in characterizing enzymes' chemical reactions. Thus, in order to avoid to get trapped into local minima, side chains are ignored and enzymes are represented exclusively through their `backbone' atoms that are carbon, nitrogen and calcium. 

Additionally, we note that enzymes do not possess any absolute spatial orientation. Unlike objects such as chairs or boats that appear usually with a specific orientation, proteins can have any orientation in 3D conformational space, thus the Cartesian coordinates defined by the model stored in PDB represent only a frozen in space and time snapshot of an overall highly dynamic structural diversity. The orientation is irrelevant to the properties of the protein. This observation underlines the need of either a rotation invariant representation or of a convention that makes output structures comparable one to another based on the definition of an intrinsic coordinate system. We define as origin of this intrinsic coordinate system the consensus barycenter of the protein as it is defined by taking into account only the four atoms of the backbone for each residue, and as axes the principal directions of each enzyme calculated by principal component analysis. Each structure is rotated around its center and the three principal directions of the enzyme aligned with the three axes of the Cartesian coordinate system.
Instead of defining a common reference frame and aligning the objects before building the prediction model, other works \cite{Brock2016} \cite{Hegde2016} \cite{Maturana2015} applied rotations around relevant axes for data augmentation. We decided to spatially normalize the data instead of arbitrarily augmenting them (by random rotations) because the number of samples is already big enough and an adequate sampling of all possible orientations would lead to an extremely large dataset difficult to handle. However, similarly to these works the definition of orientation includes uncertainties about the direction (left-right, bottom-up), which we tackled also by data augmentation.

\begin{figure}[!ht]
\centering
\includegraphics[width=0.25\textwidth]{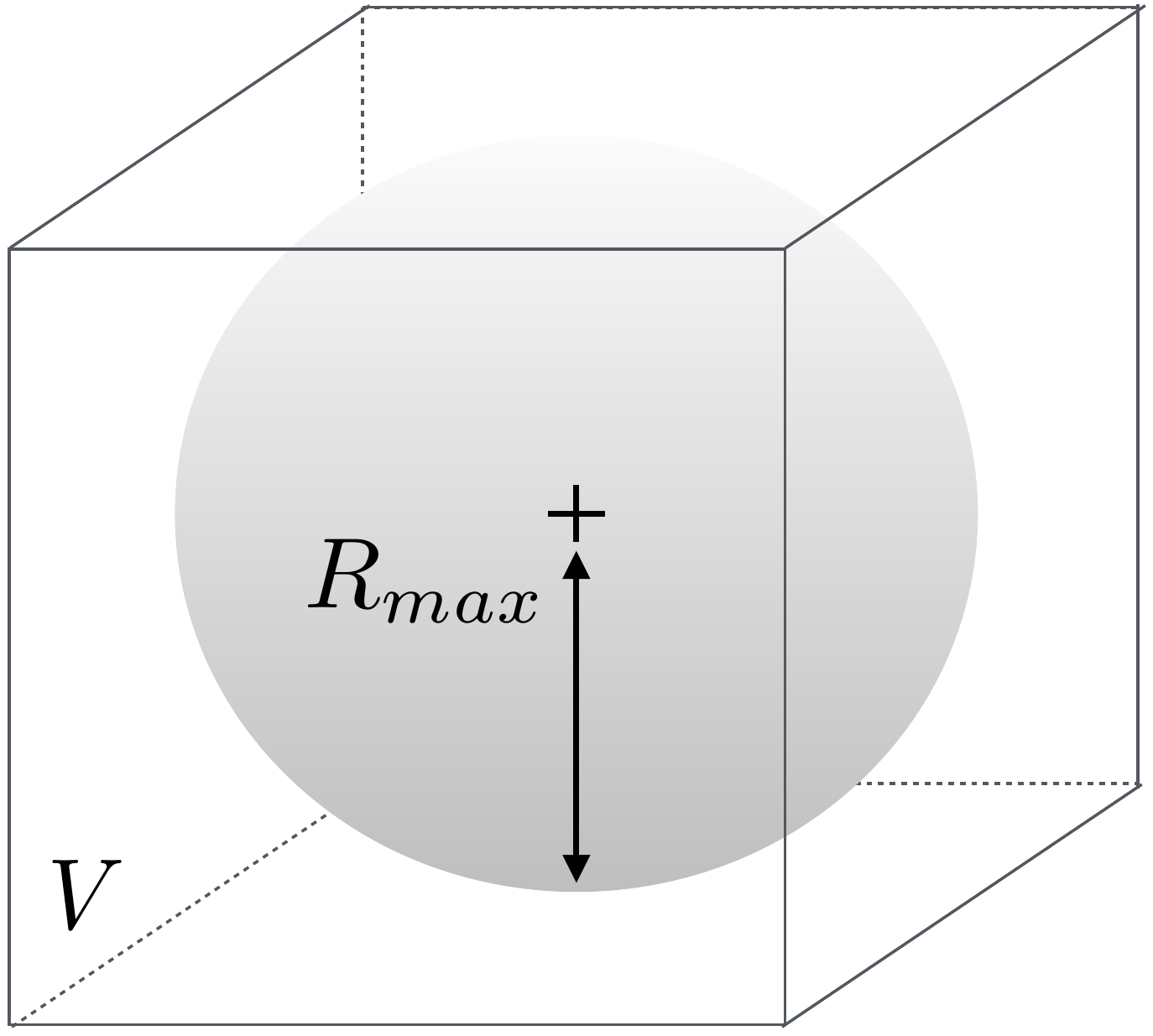}
\caption{Illustration of the meaning of $R_{max}$ with respect to volume $V$}
\label{fig:R_max}
\end{figure}

Another critical part of the process is to determine \textit{how} the enzymes should be fit in their volumes, as those can be of all types of shapes and sizes. Should we scale enzymes separately to make them fit in their respective volumes? Or on the contrary, should we scale all enzymes in a uniform manner? 
We choose to select the second option because of two reasons. First, doing otherwise would lead enzymes to be represented at different resolutions. Second, biological considerations invite us to make the convolutional network aware of the size difference between samples, as those may be an implicit feature regarding class determination.
We already know that our source files provide the coordinates of the enzymes at a same scale. After all, proteins are comprised of various combinations of equally sized amino acids. This scaling issue is therefore equivalent to determining a maximum radius $R_{max}$ so that the atom occupancy information contained in the sphere centered on the barycenter of the enzyme and of radius $R_{max}$ fits into $V$. This situation is illustrated in Figure~\ref{fig:R_max}.

\noindent $R_{max}$ has to be large enough so that a sufficient number of enzymes fit the most of their volume inside $V$. Conversely, it also has to be small enough so that most enzymes are represented at a satisfactory resolution.
 \\ As a result, a homothetic transformation with center $S$ and ratio $\lambda$ defined by
 \begin{equation}
S \text{ center of $V$} \qquad \text{and} \qquad \lambda = \left \lfloor \frac{l}{2}-1 \right \rfloor \times \frac{1}{R_{max}}
 \end{equation}
\noindent is performed on all enzymes to scale them to the desired size.

When $l$ is low, the grid is coarse enough so that the voxels of the structure have a contiguous shape. Conversely, big volumes tend to separate voxels, which engender `holes'. In that case, consecutive backbone atoms $(\vec{A_i}, \vec{A_{i+1}})$ are interpolated by $p$ regularly spaced new points computed by
\begin{equation}
\frac{(p-k+1) \times \vec{A_i} + k \times \vec{A_{i+1}}}{p+1}
\end{equation}

\noindent where $k$ varies from 1 to $p$. The latter parameter is determined empirically beforehand on enzymes of the training set. 

A last preprocessing step is to remove potential outliers from the volume. This is done by eliminating voxels that do not have any immediate neighbor.

\begin{figure}[!ht]
\begin{subfigure}{.33\textwidth}
  \centering
  \includegraphics[width=0.8\linewidth]{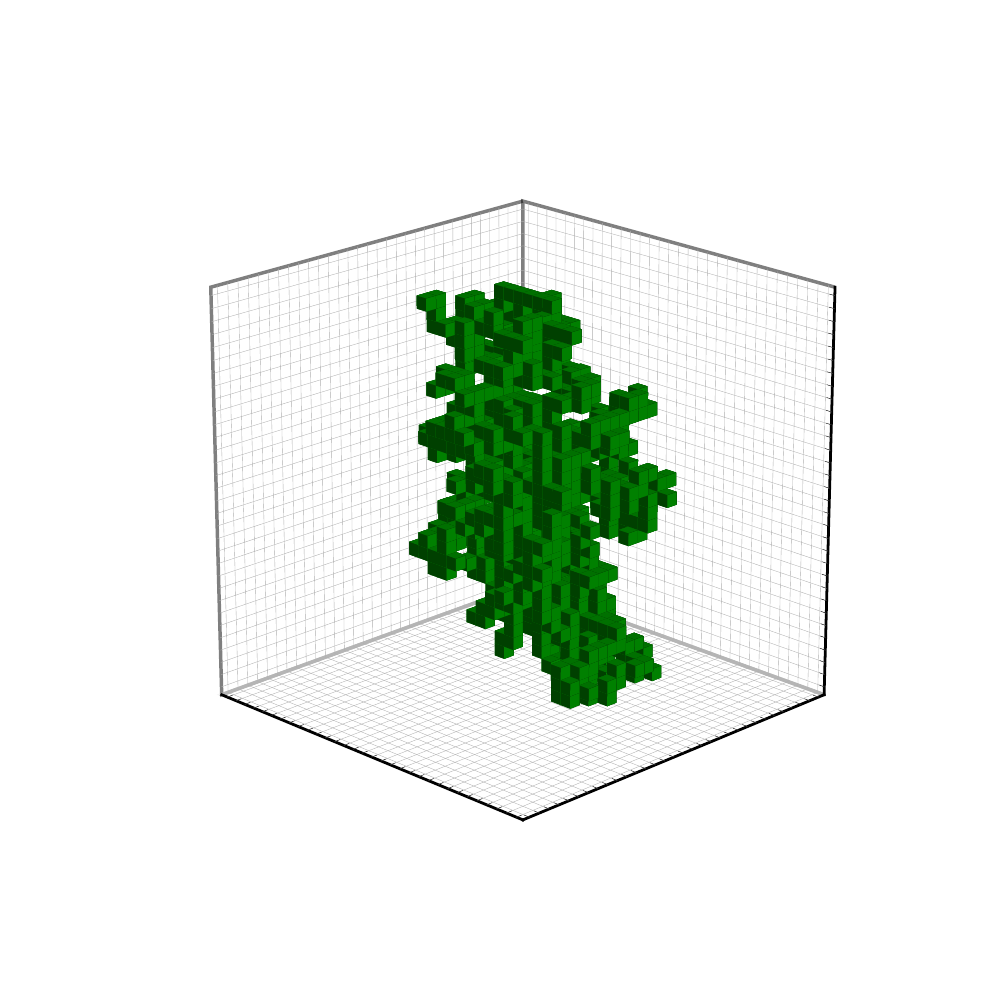}
  \caption{$l = 32$}
\end{subfigure}%
\begin{subfigure}{.33\textwidth}
  \centering
  \includegraphics[width=0.8\linewidth]{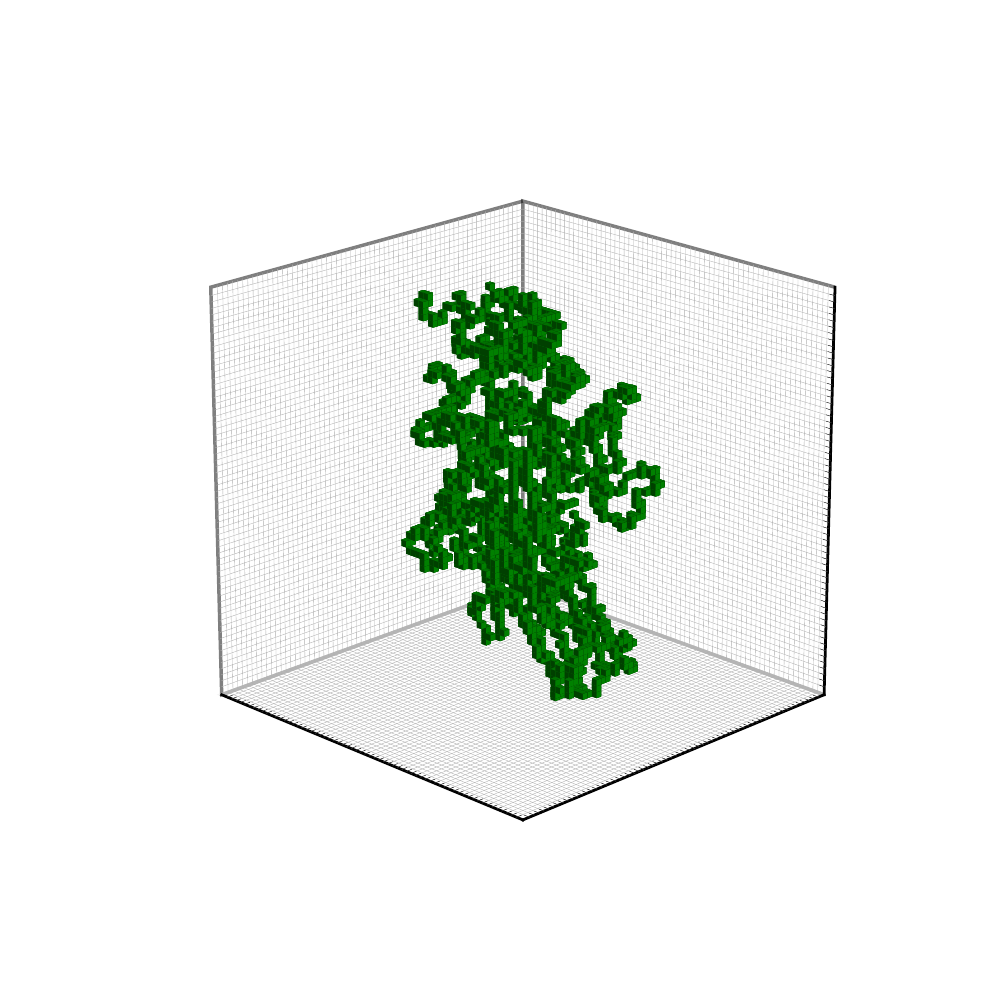}
  \caption{$l = 64$}
\end{subfigure}
\begin{subfigure}{.33\textwidth}
  \centering
  \includegraphics[width=0.8\linewidth]{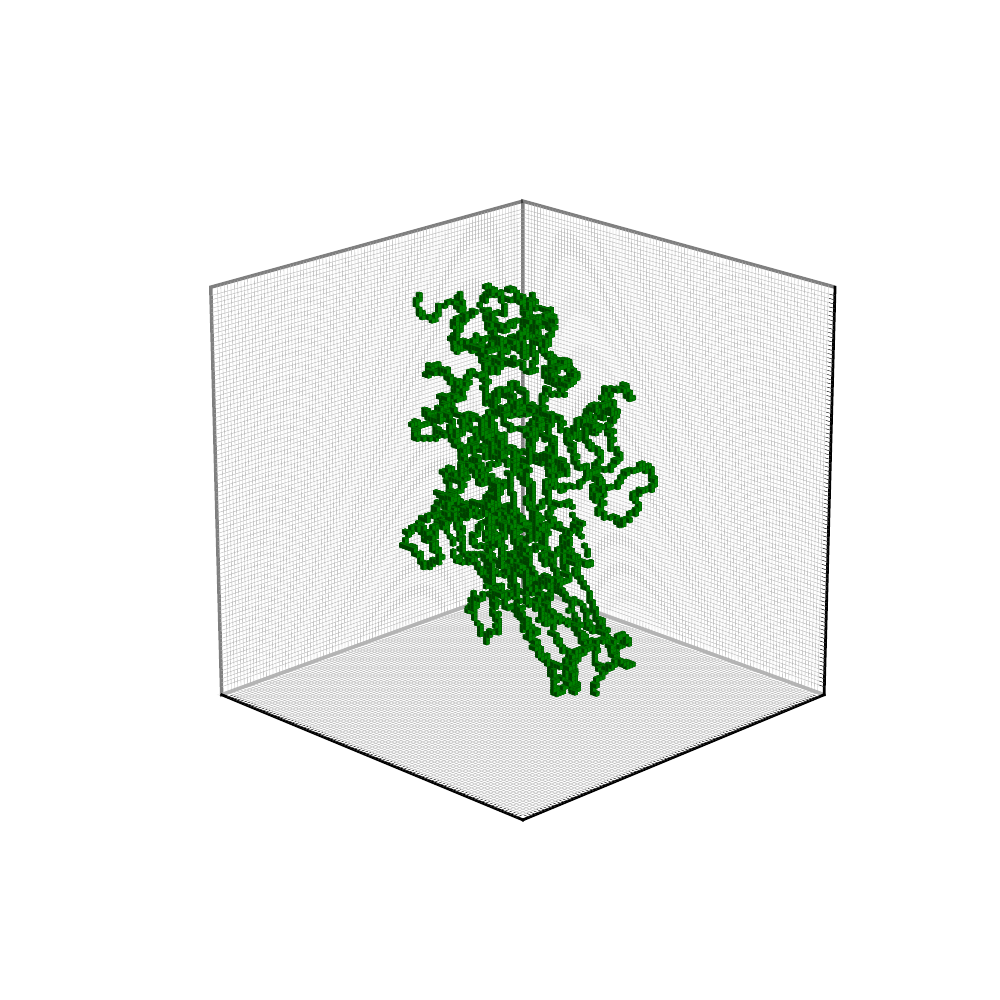}
  \caption{$l = 96$}
\end{subfigure}
\caption{Illustration of enzyme 2Q3Z for different sizes of grid}
\label{fig:grid_sizes}
\end{figure}

\noindent The illustration of the output volume obtained for different grid sizes has been provided in Figure~\ref{fig:grid_sizes}.

\subsection*{Data augmentation}
As noted earlier, differences in spatial orientation had to be considered so that volumes can be comparable one to another.

Keeping the orientation constant, we perform data augmentation by applying transformations that preserve the principal components along the three axes, \textit{i.e.} flips and combination of flips. The number of possible transformations applicable to each protein is $2^3 - 1$.

\subsection*{Architecture}
\begin{figure}[!ht]
\centering
\includegraphics[width=\textwidth]{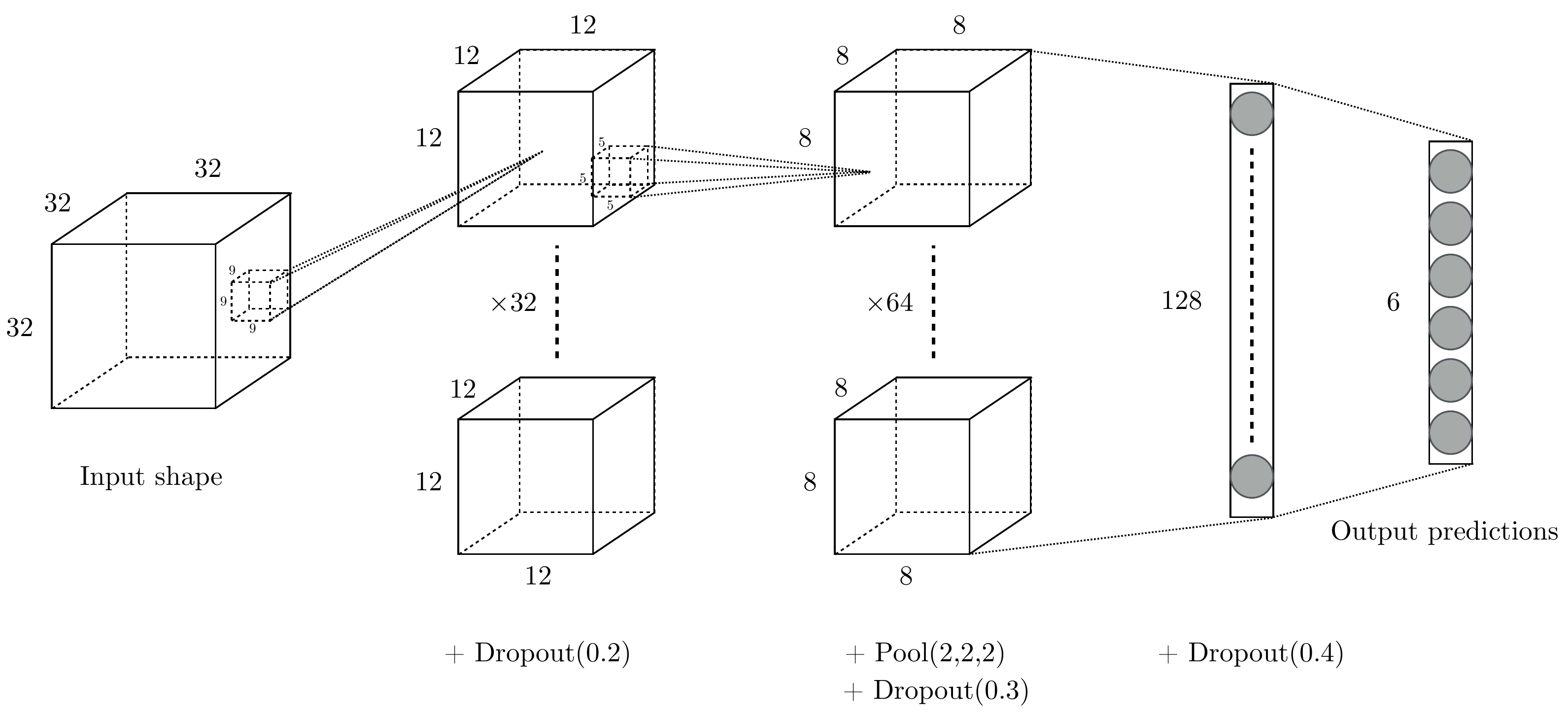}
\caption{Drawing of the architecture selected for our experiments}
\label{fig:architecture}
\end{figure}

We considered shallow architectures in order to develop a framework that can be trained with common computational means. A grid search of configurations has been conducted on the training set and led us to consider the 2-layer architecture presented in Figure~\ref{fig:architecture}: input volumes of size $32 \times 32 \times 32$ first go through a convolutional layer of 32 filters of size $9 \times 9 \times 9$ with stride 2. Then, a second convolutional layer of 64 filters of size $5 \times 5 \times 5$ with stride 1 is used, followed by a max-pooling layer of size $2 \times 2 \times 2$ with stride 2. Finally, there are two fully connected layers of 128 and 6 (the number of classes) hidden units respectively, concluded by a softmax layer that outputs class probabilities.

Several components of the VoxNet architecture \cite{Maturana2015} have also been investigated in our network. Leaky ReLU with parameter $\alpha = 0.1$ is used as activation layer after each convolutional layer. The L2 regularization technique of strength $\lambda = 0.001$ is applied on the network's layers. Overfitting is also tackled using dropout throughout the network.

All in all, this model contains $804,614$ distinct parameters (including biases), which is approximately 13\% less than for the VoxNet architecture \cite{Maturana2015}.

\noindent Additionally, the Adam optimizer has been chosen for its computational efficiency and low need of hyper-parameter tuning \cite{Kingma2015}.

The categorical cross entropy is used as loss function, since the latter can be adapted to multiclass classification. Two approaches are considered for computing this loss. The first one assesses misclassification error irrespectively of individual class sizes, whereas the second one takes class imbalance into account and uses customized penalization weights for each class. In this second approach, the aim is to compensate the lack of training samples in under-represented classes by providing a greater penalization to the loss in the event of misclassification than for larger classes.
\\The cross entropy loss is given by
\begin{equation}
\mathcal{L} = -\sum_{x\in\mbox{{\tiny {set}}}}\sum_{i \, = \, 1}^6 w_i \cdot \delta_{x,i} \cdot \log(\widehat{p}_{x,i})
\end{equation}
where $\widehat{p}_{x,i}$ is the predicted probability of enzyme $x$ belonging to class EC$i$, $\delta_{x,i}$ the quantity equal to 1 only if enzyme $x$ belongs to class EC$i$, and $w_{i}$ the weight associated to class EC$i$. For the first approach, all weights $w_i$ are taken equal to 1.
For the weighted approach, we chose weights according to the formula
\begin{equation}
w_i = \frac{\displaystyle \max_{j \in [\![1;6]\!]} \#(\text{EC$j$ training enzymes})}{\#(\text{EC$i$ training enzymes})}
\end{equation}
which increases the contribution of the under-represented classes by an amount inversely proportional to their size and in respect to the largest class.

\subsection*{Metrics}
Among the various multi-class metrics that have been studied in \cite{Sokolova2009}, we selected the most representative to assess the model's performance. The metrics are based on the confusion matrix whose elements $C(i,j)$ with $i,j\in[\![1,6]\!]$ indicate the number of enzymes that belong to class EC$i$ and are predicted as belonging to class EC$j$:

\begin{itemize}
\item \textit{Accuracy} which captures the average per-class effectiveness of the classifier:

\[\mbox{Accuracy} = \frac{\displaystyle\sum_{i=1}^6C(i,i)}{\displaystyle\sum_{i,j=1}^6C(i,j)}\]

\item \textit{Precision, recall} and \textit{F1} score which are calculated per class EC$i$:

\[\mbox{Precision}_{{\tiny\mbox{EC}i}}=\frac{C(i,i)}{\displaystyle\sum_{j=1}^6C(j,i)}\hspace{2\baselineskip}\mbox{Recall}_{{\tiny\mbox{EC}i}}=\frac{C(i,i)}{\displaystyle\sum_{j=1}^6C(i,j)}\]

\[\mbox{F1}_{{\tiny\mbox{EC}i}}=2\times\frac{\mbox{Precision}_{{\tiny\mbox{EC}i}}\times\mbox{Recall}_{{\tiny\mbox{EC}i}}}{\mbox{Precision}_{{\tiny\mbox{EC}i}}+\mbox{Recall}_{{\tiny\mbox{EC}i}}}\]

\noindent For each class, precision gives us an idea of the proportion of correctly classified enzymes among enzymes that have been classified in that class, while recall highlights the proportion of correctly classified enzymes among enzymes that actually belong to that class.

\item \textit{Macro precision, recall} and \textit{F1} score which express average performance over the 6 enzyme classes:

\[\mbox{Precision}_M=\frac{1}{6}\sum_{i=1}^6\mbox{Precision}_{{\tiny\mbox{EC}i}}\hspace{2\baselineskip}\mbox{Recall}_M=\frac{1}{6}\sum_{i=1}^6\mbox{Recall}_{{\tiny\mbox{EC}i}}\]

\[\mbox{F1}_M=2\times\frac{\mbox{Precision}_M\times\mbox{Recall}_M}{\mbox{Precision}_M+\mbox{Recall}_M}\]

\end{itemize}

\subsection*{Final decision rule}
Several approaches are considered for determining an enzyme's final class.
\begin{itemize}
\item The first strategy is to make the prediction based on the model of the 3D shape without any transformation.
\item In the second approach, the $2^3-1$ possible combinations of flipped volumes are generated and introduced to the classifier. The final prediction is either the class of maximum total probability (\textit{probability}-based decision) or the class selected by majority voting (\textit{class}-based decision) 
\item The third strategy is also based on fusion of decisions produced for each flipped volume, but this one weights each decision by a different coefficient, such as $\frac{1}{\delta_x + \delta_y + \delta_z + 1}$, which highlights transformations with the least number of flips.
\end{itemize} 

\subsection*{Hyperparameter selection}
\subsubsection*{Radius $R_{max}$ and interpolation parameter $p$}
\begin{figure}[!ht]
\begin{subfigure}{.5\textwidth}
  \centering
  \includegraphics[width=\linewidth]{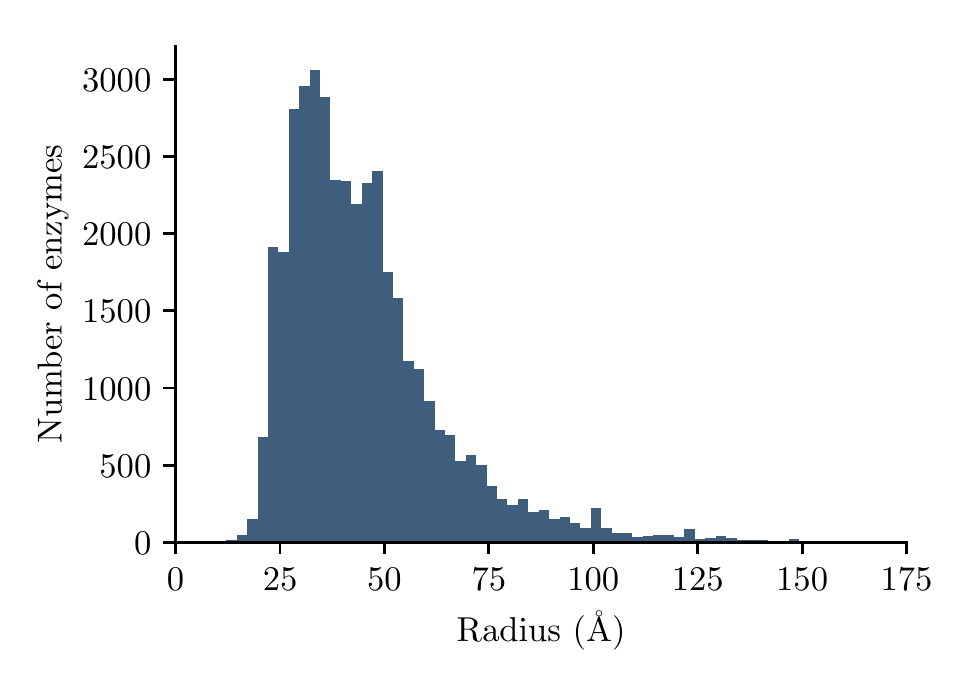}
  \caption{Histogram of the radii of training enzymes}
  \label{fig:analysis_rmax_1}
\end{subfigure}
\begin{subfigure}{.5\textwidth}
  \centering
  \includegraphics[width=\linewidth]{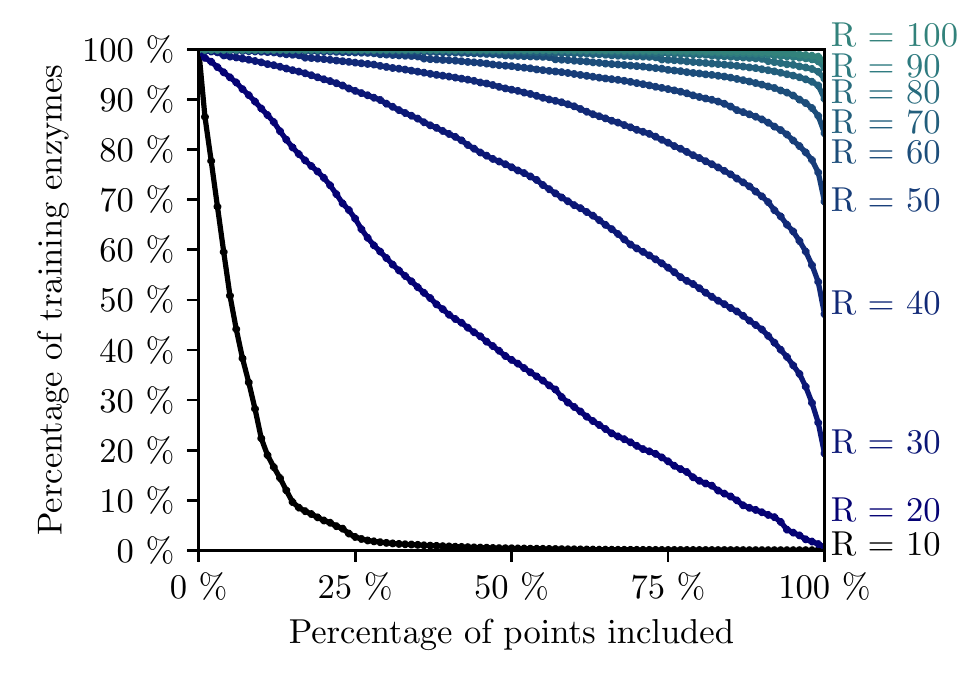}
  \caption{Level of information conveyed for each radius}
  \label{fig:analysis_rmax_2}
\end{subfigure}
\caption{Analysis of the radii distribution on the training set}
\label{fig:analysis_rmax}
\end{figure}

A crucial point in the presented approach is to make a cogent selection of $R_{max}$. As previously discussed, this parameter controls the trade-off between the level of information ($l$) retained in each volume and the resolution with which they are conveyed ($R_{max})$.
Figure~\ref{fig:analysis_rmax} shows the analysis of the dataset from these two perspectives. The graph on the left helps us assess the minimum radius for which a decent amount of enzymes will be totally included in the volume, whereas the graph on the right highlights the quantity of information retained by each radius.

\noindent In details, from Figure~\ref{fig:analysis_rmax_1} we can see that the majority of enzymes can fit in a sphere with radius between 25 and 75 \AA, with a peak around 35 \AA. In Figure~\ref{fig:analysis_rmax_2}, each point $(x,y)$ on a curve of radius $R$ shows the percentage $y$ of training enzymes that have at least $x$ points included within radius $R$ around their respective enzyme barycenter. Based on both graphs, we select a value of $R_{max} = 40$, that is big enough to capture more than half of enzymes in $V$, but small enough so that the smallest enzymes have radii of at least half of $R_{max}$.

Empirical observations show that a grid of $l = 32$ does not require any atom interpolation. Therefore, $p$ is set to zero for our computations. For denser grids, \textit{e.g.} with grid size $l = 64$ or 96, appropriate $p$ values were $p = 5$ and $p = 9$, respectively.

\subsubsection*{Random selection of transformations and samples}
Regarding data augmentation, the probability of enzyme flip along each axis has been set to $\frac{2}{10}$. That way for each enzyme, higher numbers of flips have a lower probability of happening. Also for each pass, approximately half randomly selected enzymes are to be augmented by flips (or a combination of flips). This process is useful, as it will help us obtaining a robust classifier.

\section*{Results}
\begin{figure}[!ht]
\begin{subfigure}{.5\textwidth}
  \centering
  \includegraphics[width=\linewidth]{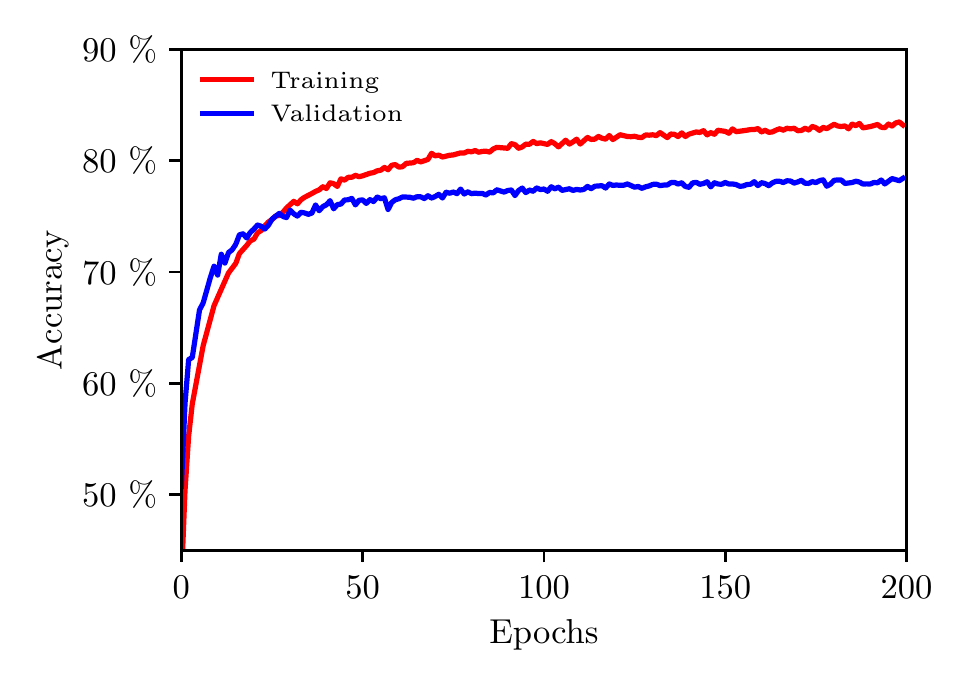}
  \caption{Uniform weights}
\end{subfigure}
\begin{subfigure}{.5\textwidth}
  \centering
  \includegraphics[width=\linewidth]{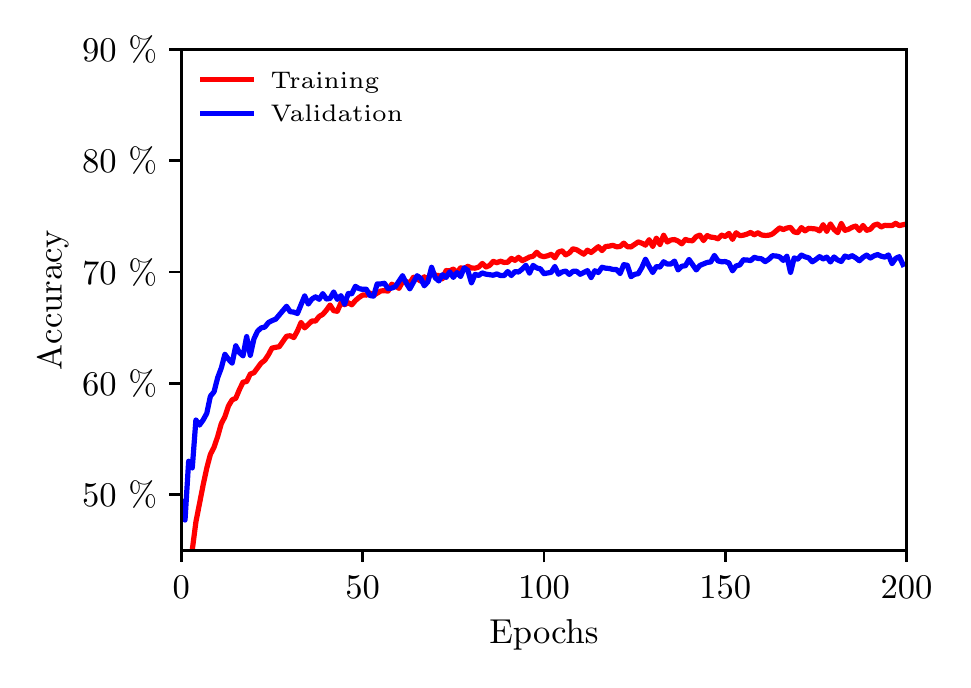}
  \caption{Adjusted weights}
\end{subfigure}
\caption{Evolution of various indicators during training process}
\label{fig:plot_accuracy_epochs}
\end{figure}

We trained the model with and without weights adjustment in the loss function. The evolution of performance with increasing number of epochs is shown in Figure~\ref{fig:plot_accuracy_epochs}. It can be seen that the training accuracy is almost stabilized after 200 epochs. Thus the learnt weights at 200 epochs were used for the final prediction model during testing. A higher number of epochs is possible to overfit the data.

The testing results are shown in Table~\ref{table:testing_results}.

\renewcommand{\arraystretch}{1.1}
\setlength{\tabcolsep}{4.0pt}
\begin{table}[!h]
\begin{center}
\resizebox{\columnwidth}{!}{
\begin{tabular}{|c|c|c|ccccc|ccccc|}
\hline
\multicolumn{3}{|c|}{Weights}&\multicolumn{5}{|c|}{Uniform}&\multicolumn{5}{|c|}{Adapted}\\
\cline{1-13}
\multicolumn{3}{|c|}{Decision}&\multicolumn{1}{|c|}{\multirow{2}{*}{None}}&\multicolumn{2}{|c|}{Probabilities}&\multicolumn{2}{|c|}{Classes}&\multicolumn{1}{|c|}{\multirow{2}{*}{None}}&\multicolumn{2}{|c|}{Probabilities}&\multicolumn{2}{|c|}{Classes}\\
\cline{1-3}\cline{5-8}\cline{10-13}
\multicolumn{3}{|c|}{Augmentation}&&\multicolumn{1}{|c|}{Flips}&\multicolumn{1}{|c|}{W. flips}&\multicolumn{1}{|c|}{Flips}&\multicolumn{1}{|c|}{W. flips}&&\multicolumn{1}{|c|}{Flips}&\multicolumn{1}{|c|}{W. flips}&\multicolumn{1}{|c|}{Flips}&\multicolumn{1}{|c|}{W. flips}\\
\hline
\multicolumn{3}{|c|}{\multirow{1}{*}{Accuracy}}&\textbf{0.784}&0.761&0.781&0.756&0.775&0.707&0.714&0.729&0.708&0.722\\
\hline
\multicolumn{1}{|c|}{\multirow{7}{*}{Precision}}&\multicolumn{1}{|c|}{\multirow{6}{*}{EC}}&1&0.846&0.908&\textbf{0.910}&0.863&0.892&0.766&0.791&0.800&0.719&0.773\\
\cline{3-3}
&&2&0.757&0.735&0.750&0.715&0.752&0.773&0.774&\textbf{0.786}&0.753&0.777\\
\cline{3-3}
&&3&0.755&0.705&0.729&0.720&0.725&\textbf{0.791}&0.755&0.769&0.764&0.778\\
\cline{3-3}
&&4&0.928&0.969&\textbf{0.971}&0.953&0.958&0.693&0.825&0.829&0.839&0.789\\
\cline{3-3}
&&5&0.918&0.972&0.970&\textbf{0.975}&0.952&0.467&0.571&0.582&0.579&0.566\\
\cline{3-3}
&&6&0.807&0.951&0.968&\textbf{0.973}&0.942&0.271&0.286&0.298&0.301&0.295\\
\cline{2-13}
&\multicolumn{2}{|c|}{Macro}&0.835&0.873&\textbf{0.883}&0.866&0.870&0.627&0.667&0.677&0.659&0.663\\
\cline{1-13}
\hline
\multicolumn{1}{|c|}{\multirow{7}{*}{Recall}}&\multicolumn{1}{|c|}{\multirow{6}{*}{EC}}&1&\textbf{0.765}&0.710&0.740&0.724&0.737&0.743&0.719&0.744&0.740&0.743\\
\cline{3-3}
&&2&0.798&0.787&\textbf{0.807}&0.805&0.800&0.654&0.644&0.670&0.653&0.662\\
\cline{3-3}
&&3&0.872&0.897&\textbf{0.901}&0.870&0.896&0.737&0.774&0.778&0.751&0.767\\
\cline{3-3}
&&4&0.659&0.541&0.586&0.516&0.584&\textbf{0.717}&0.685&0.704&0.676&0.703\\
\cline{3-3}
&&5&0.546&0.426&0.459&0.408&0.454&0.680&0.687&0.687&0.662&\textbf{0.695}\\
\cline{3-3}
&&6&0.407&0.243&0.310&0.229&0.306&0.655&\textbf{0.730}&0.721&0.669&0.709\\
\cline{2-13}
&\multicolumn{2}{|c|}{Macro}&0.675&0.601&0.634&0.592&0.629&0.698&0.706&\textbf{0.717}&0.692&0.713\\
\cline{1-13}
\hline
\multicolumn{1}{|c|}{\multirow{7}{*}{F1}}&\multicolumn{1}{|c|}{\multirow{6}{*}{EC}}&1&0.804&0.797&\textbf{0.816}&0.787&0.807&0.754&0.753&0.771&0.729&0.758\\
\cline{3-3}
&&2&0.777&0.760&\textbf{0.778}&0.757&0.775&0.709&0.703&0.724&0.699&0.715\\
\cline{3-3}
&&3&\textbf{0.809}&0.790&0.806&0.788&0.801&0.763&0.764&0.774&0.757&0.772\\
\cline{3-3}
&&4&\textbf{0.770}&0.694&0.731&0.670&0.726&0.705&0.749&0.761&0.749&0.744\\
\cline{3-3}
&&5&\textbf{0.685}&0.592&0.623&0.575&0.614&0.554&0.623&0.630&0.618&0.624\\
\cline{3-3}
&&6&\textbf{0.541}&0.387&0.469&0.370&0.462&0.383&0.411&0.422&0.415&0.417\\
\cline{2-13}
&\multicolumn{2}{|c|}{Macro}&\textbf{0.746}&0.712&0.738&0.703&0.730&0.660&0.686&0.697&0.675&0.687\\
\hline
\end{tabular}
}
\end{center}
\caption{Testing results}
\label{table:testing_results}
\end{table}

\noindent Overall, the model that performs best in terms of both macro accuracy (78.4\%) and macro F1 (74.6\%) is the one with uniform weights using no data augmentation.

Precision per class on under-represented classes (EC4, 5, 6) is far better on uniformly weighted models compared to adapted models, with best scores being 97.1\%, 97.5\% and 97.3\%, versus 83.9\%, 58.2\% and 30.1\% respectively. Over-represented classes (EC2, 3) have roughly the same performance in those two types of model.

It is worth noting that by augmenting the data, the precision per class is noticeably improved on under-represented classes (EC4, 5, 6) in both uniform and adapted models. In fact, in the uniformly weighted framework, the best data augmented models achieve 97.1\%, 97.5\% and 97.3\% versus 92.8\% 91.8\% and 80.7\% for the non-augmented one for EC4, 5, 6 respectively. Similarly, in the adapted framework, they achieve 83.9\%, 58.2\% and 30.1\% versus 69.3\%, 46.7\% and 27.1\% for EC4, 5, 6 respectively.

Recall per class on over-represented classes (EC2, 3) are best using uniformly weighted models, with 80.7\% and 90.1\% respectively. It is worth noting that adapted models outperform uniform ones on under-represented classes (EC4, 5, 6) with 71.7\%, 69.5\%, 73.0\% recall per class respectively, compared to 65.9\%, 54.6\% and 40.7\% respectively on uniform models.

Uniformly weighted models perform best in terms of F1 scores, ranging from 54.1\% for the smallest class (EC6) to 81.6\% for EC1, with a macro F1 of 74.6\%. 

\subsection*{Computation time}
Our architecture was implemented on Python 3 using Keras \cite{Chollet2015} on top of a GPU-enabled version of TensorFlow \cite{TensorFlow2015}. Enzyme information has been extracted using the open-source module BioPython, a fast and easy-to-use tool presented in \cite{Cock2009}. The average prediction time of a single enzyme was about 6 ms without flips, or 50 ms with flips, on an Intel i7 6700K machine with 32 GB of RAM and a GTX 1080 graphics card.

\section*{Discussion}
The general trend is that uniformly weighted models perform well in terms of macro accuracy, macro precision in respect to macro F1
while adapted models are slightly better in macro recall.
More particularly, on under-represented classes (EC4, 5, 6), models perform better in terms of precision per class when using flip data augmentation, which means that using data augmentation increases reliability on predictions. This can be explained by the fact that the classifier has more examples at hand, which makes its predictions more robust.
Interestingly, on under-represented classes, uniform models have the highest precision per class and the lowest recall per class while adapted models are exactly the opposite: they have the lowest precision per class and the highest recall per class. The interpretation of this clear trend is that enzymes coming from under-represented classes are not always recognized by uniform classifiers as they are biased towards enzymes from over-represented class because the classifier does not correct for class imbalance. On the contrary, enzymes from under-represented classes are well recognized by adapted classifiers as they account for class imbalance, but this comes at a price: by predicting more enzymes as being in those under-represented classes (false positives), the classifier tends to make a lot more mistakes which leads to a low precision per class.

\subsection*{Further improvements of the method}
In a parallel we investigated possible improvements in the architecture. We introduced batch normalization \cite{Ioffe2015} at several positions of the network and repeated the same experiments as before. Those were placed after every activation layer, as suggested in \cite{Mishkin2016}. Also, Leaky ReLU activation layers were replaced by PReLU \cite{He2015} layers which enable the network to learn adaptively the Leaky ReLU's best $\alpha$ parameter. Although these changes led the network to converge faster to a stable optimum, the final performance increase was only marginal and came at a higher computational cost.

\begin{figure}[!ht]
\begin{subfigure}{.5\textwidth}
  \centering
  \includegraphics[width=0.528\linewidth]{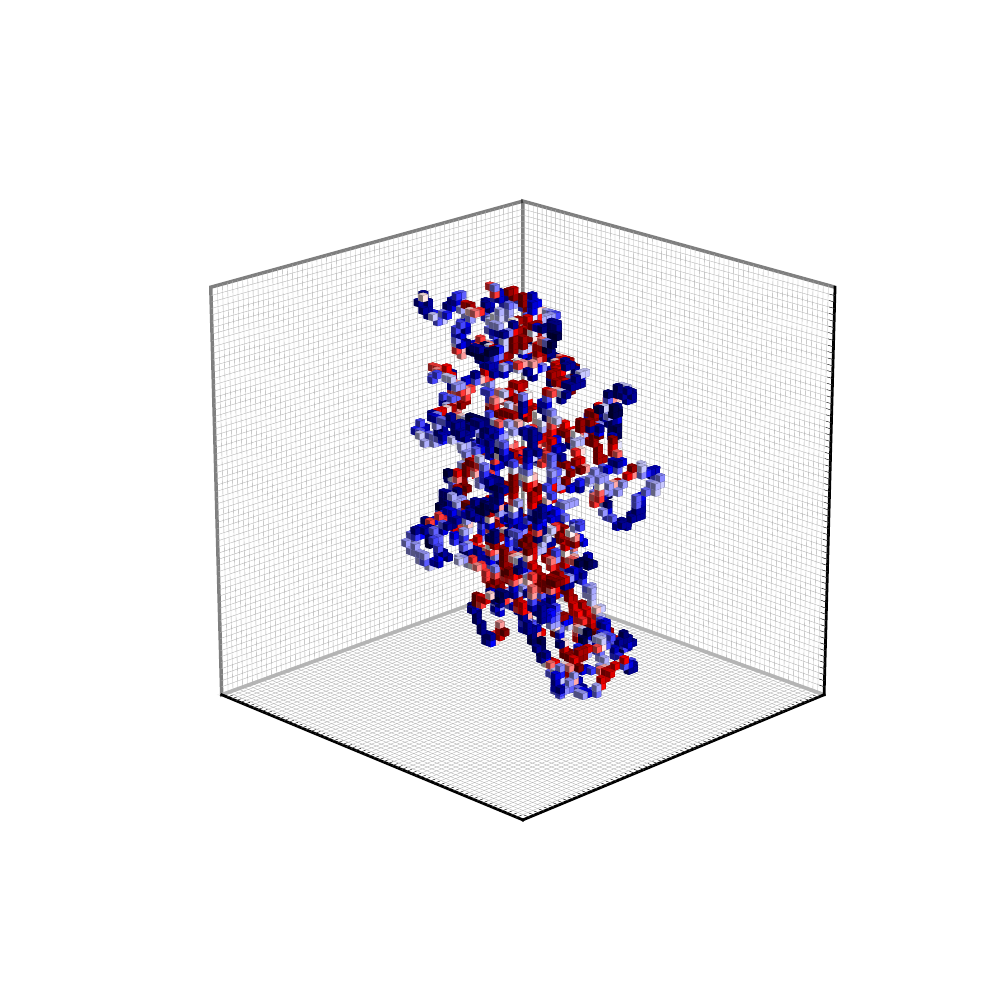}
  \caption{Hydropathy}
  \label{fig:sfig2}
\end{subfigure}
\begin{subfigure}{.5\textwidth}
  \centering
  \includegraphics[width=0.528\linewidth]{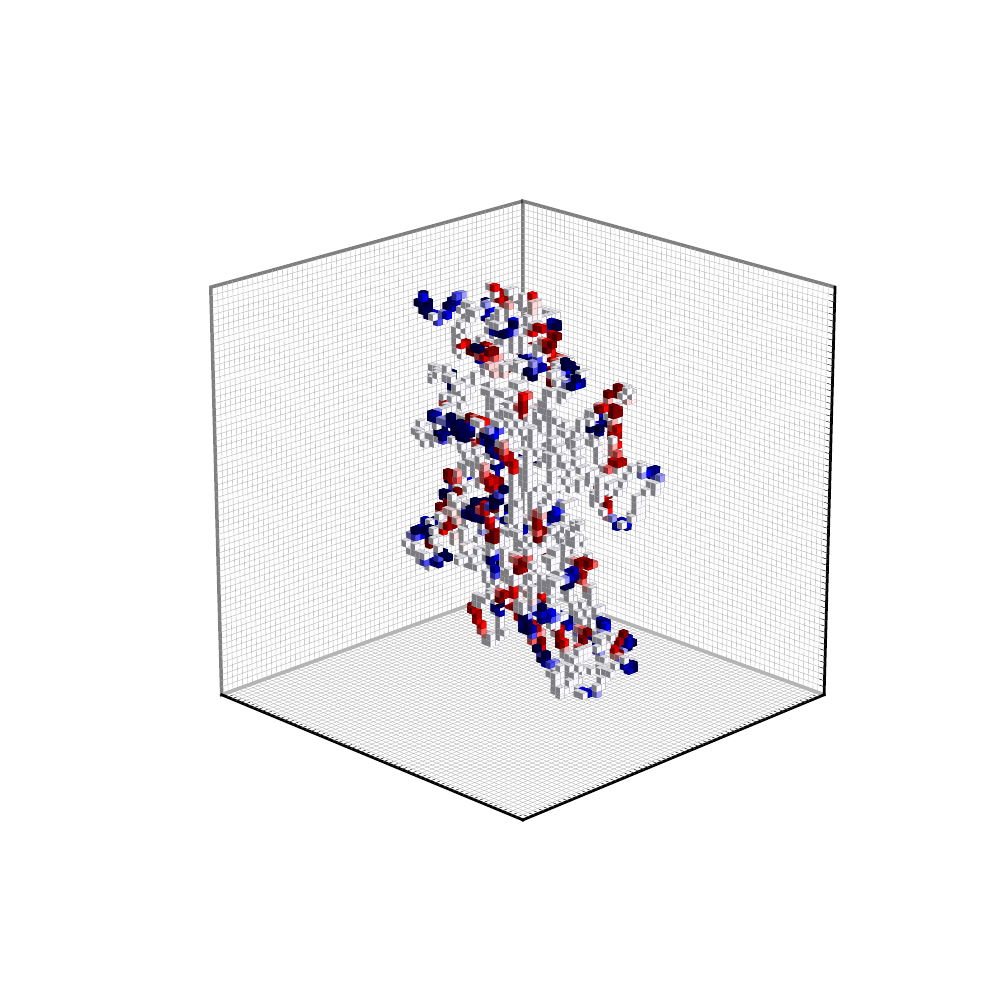}
  \caption{Charge}
  \label{fig:sfig2}
\end{subfigure}
\caption{Illustration of several attributes incorporated into the shape model for enzyme 2Q3Z. Each attribute is illustrated in color for better visualization, but it actually corresponds to a single channel.}
\label{fig:additional_features}
\end{figure}

In the following, we considered the transition from a binary representation capturing only shape information to "gray-level" representation capturing also information content. Several biological indicators such as the hydropathy index \cite{Kyte1982}, or isoelectric points \cite{Wade2002} can be used to better describe local properties of amino-acids that are the building blocks of the protein structure. From the perspective of computational analysis, these attributes can be incorporated into the representation model by attaching to the  shape also appearance information.  Figure~\ref{fig:additional_features} illustrates this idea showing the volumetric representation of two different attributes (hydropathy on the left and charge on the right). The different attributes, if up to 3, can also be merged into a single structure visualized in color (RGB) scale. 
Different approaches to handle multi-channel volumetric data (images) by CNNs have recently been presented \cite{Zhang2017} \cite{Cao2016} \cite{Kamnitsas2017} \cite{Nie2016}. The literature however on the use of deep networks for 3D shapes with multi-channel appearance is limited. 
As preliminary analysis, we applied EnzyNet without modification on the architecture, but with re-training for the adjustments of the weights of the convolutional kernels. We introduced to the network either a single attribute (as a binary 3D image or gray-level 3D image), such as shape, hydropathy and isoelectric points, or a combination of these different attributes.
Two other ways of information combination were examined, either each attribute was introduced  as a different channel in the convolutional neural network, or the outputs of the single channel networks were combined through a fusion rule. However, all methods showed an increase of the order of magnitude of one percent. Further investigation will be required to appropriately harness this added information.

Furthermore, in the current study  a relatively small grid size was selected due to computational limitations. However, as can be seen in Figure~\ref{fig:grid_sizes} the details of the spatial structure of proteins can be better differentiated for higher values of $l$. The next step of this work would be to adapt the current architecture accordingly into a similar network that processes higher-resolution volumes. This could enable the network to capture more subtle features and potentially boost the performance of the classifier.

Wei et al. \cite{Wei2015} present a method that enables 3D convolutional neural nets to deal with multi-label classification problems. This approach is interesting for us as it allows the extension of our method to the classification of multi-label enzymes. The obtained performance could subsequently be compared to previous work \cite{Amidi2017}.

Other approaches based on the 3D representation of enzymes are also possible. Brock et al. \cite{Brock2016} performed very well on the ModelNet dataset using generative and discriminative modeling. Their voxel-based autoencoder is helpful for assessing the key features that are correctly learned from the 3D shapes. We could elaborate this information in order to identify significant features in a pre-training phase aiming to obtain better prediction performance.


\bibliographystyle{plain}
\bibliography{bibli}

\end{document}